\documentclass[published]{JHEP3}

\title{The cosmological constant and oscillating metrics}

\author{Hael Collins\thanks{Current address: Department of Physics, Carnegie
Mellon University, Pittsburgh, PA 15213, USA}~ and Bob Holdom\\
Department of Physics, University of Toronto\\
Toronto, Ontario M5S 1A7, Canada\\ 
E-mail: \email{hael@cmuhep2.phys.cmu.edu},
\email{bob.holdom@utoronto.ca}}

\abstract{The presence of a cosmological constant, $\Lambda$, in an
action with higher powers of the curvature can produce rapidly
oscillating metrics. We develop a perturbative approach for generating
periodic solutions to the non-linear field equations for such actions
based on a small amplitude expansion.  We find that these oscillations
have an amplitude proportional to $\sqrt{\Lambda}$ and a frequency of
order the Planck mass.  In a $4+1$-dimensional scenario, a family of
metrics exists that are periodic in the extra dimension and are
parameterized by an effective four-dimensional cosmological constant
which drives a rapid oscillation.}


\received{September 17, 2002}

\accepted{October 24, 2002}

\JHEP{10(2002)052}

\begin{document}

\section{Introduction}

Understanding the unnaturally small size of the cosmological constant
poses one of severest challenges for a theory of
gravity~\cite{weinberg}. At late times and for large distances, the
apparent size of the cosmological constant is constrained to be
extremely small in terms of the natural scale for gravity, the Planck
mass.  In contrast, no observations bound the value of the
cosmological constant during the earliest stages of the universe, when
corrections to the Einstein-Hilbert action were non-negligible, and
its presence can lead to a richer family of metrics.  Among the
solutions for a more general gravitational action, the presence of a
positive cosmological constant does not inevitably lead to a de Sitter
expansion.  Such solutions must still yield or evolve into a low
energy theory in which the \emph{effective} cosmological constant is
small to be phenomenologically acceptable.  If the characteristic
scales on which these metrics vary are of extremely high energy or
short distance, then it may be possible to integrate out such features
to arrive at a slowly varying effective theory.

To determine whether an action for gravity, generalized beyond an
Einstein-Hilbert term, admits these features --- natural
coefficients for the terms in the action and a rapid variation
--- we must first solve the highly non-linear field equations.
This task is difficult even when only the next curvature
corrections are added.  In $3+1$ dimensions, Horowitz and
Wald~\cite{horowitz} and later Starobinsky~\cite{star} discovered
oscillating solutions for actions that included quadratic
curvature terms but no cosmological constant. Numerical solutions
were found in $4+1$ dimensions~\cite{periodic} in the presence of
a cosmological constant and a scalar field, along with the
quadratic curvature terms. In this latter scenario, metrics exist
that depend periodically on the extra spatial coordinate so that
choosing the size of the extra \pagebreak[3] dimension to be the
period produces a compact extra dimension without any
fine-tunings or singularities. The parameters in the action fix
the size of the extra dimension uniquely.  However, without an
analytical approach it becomes difficult to generalize these
solutions to include an evolution in time.  Without this freedom,
it is difficult to understand how a universe starting from a more
general state can find itself in one of these configurations.

In this article, we introduce a perturbative method for finding
metrics with a periodic dependence on one or more coordinates.
The perturbative parameter in this approach corresponds to the
amplitude of the oscillation, which is generically of order
$\sqrt{\Lambda}/M$, where $\Lambda$ is the cosmological constant
and $M$ is the Planck mass. While this perturbative approach
complements the numerical analysis of the static metrics
in~\cite{periodic}, it also allows the time-dependent extension
of those metrics to be explored, which is too complicated for a
numerical analysis.  Moreover, this analytic, small amplitude
expansion allows us to extract the dependence of the properties
of the solution on the fundamental parameters in the action.  In
particular, we find the leading behavior for the amplitude and
frequency of the oscillations in terms of the cosmological
constant and coefficients of the four derivative curvature
invariants.

These oscillating solutions are useful in $4+1$ dimensions since
they provide an alternative to a purely de Sitter expansion when
the cosmological constant in the full action is not exponentially
small. In this scenario, the cosmological constant plays a dual
role --- both driving a rapid, Planck-frequency oscillation as
well as supporting a compact extra dimension. As noted
in~\cite{horowitz} such oscillations would need to be damped out
by some early stage in the universe --- only  an extremely small
amplitude is allowed today~\cite{fermions}.  The important
property of these solutions is that the relative role of the
cosmological constant in supporting the oscillations and the
compactness is not fixed by the action.  Unlike a de Sitter
universe where a natural value for the cosmological constant is
directly related to an extremely rapid expansion, in the
oscillating scenario, this relative freedom does not preclude a
large initial oscillation from subsequently relaxing, either by
particle production or by interactions with charged particles.

In section~\ref{sec:two} we introduce the small amplitude
expansion in $3+1$ and $4+1$ dimensions for metrics that only
depend on time.  This setting avoids the complications of two
variable solutions, yet illustrates some of the general
properties of the oscillating solutions. We also briefly review
the solution with a time independent metric in $4+1$ dimensions
that depends periodically on the fifth dimension.
Section~\ref{sec:three} then generalizes this compact
$4+1$-dimensional universe to include a rapidly oscillating time
dependence.  In section~\ref{sec:four}, we briefly discuss the
compatibility of these solutions with an ordinary cosmology in
the large~dimensions.

\section{Periodic metrics in the small amplitude limit}\label{sec:two}

\subsection{Oscillating metrics in $3+1$ dimensions}
Periodic metrics exist in any number of dimensions when we
generalize the standard Einstein-Hilbert action to include not
only a cosmological constant but higher derivative corrections as
well.  For simplicity, we first show the existence of an
oscillating metric in $3+1$ dimensions.  We begin with a
gravitational action with arbitrary powers of the
curvature,\footnote{Our convention for the signature of the
metric is $(-,+, \dots ,+)$ while the Riemann curvature tensor is
defined by $-R^\kappa_{\ \lambda\mu\nu} \equiv
\partial_\nu \Gamma^\kappa_{\lambda\mu} - \partial_\mu
\Gamma^\kappa_{\lambda\nu} + \Gamma^\kappa_{\rho\nu}
\Gamma^\rho_{\lambda\mu} - \Gamma^\kappa_{\rho\mu}
\Gamma^\rho_{\lambda\nu}$.  Our sign convention for $\Lambda$ has
changed from~\cite{periodic}.}
\begin{eqnarray}
S &=& M^2 \int d^4x \sqrt{-g}
\left( -2 \Lambda + R + a R^2 + b R_{\mu\nu}R^{\mu\nu}
+ c R_{\mu\nu\lambda\kappa}R^{\mu\nu\lambda\kappa} + \cdots \right) \\
&& +\,  M^2 \int d^4x \sqrt{-g} \mathcal{L}\,,
\nonumber
\end{eqnarray}
and consider solutions of the form,
\begin{equation}
ds^2 = - dt^2 + e^{B(t)} g_{ij} dx^i dx^j\,.
\label{Bmetric}
\end{equation}
We assume that $\mathcal{L}$ contains a free massless scalar
field, $\phi(t)$, and in particular contributes to the
energy-momentum tensor as a perfect fluid with density, $\rho$,
and pressure, $p$, of $\rho = p = \frac{1}{4}\dot\phi^2+C$.  $C$
is some additional constant contribution, perhaps arising as a
self-consistent Casimir effect.

We shall show the existence of solutions where the function
$B(t)$ is a smooth, periodic function of time.  When the
amplitude, $\epsilon$, of these oscillations is small, this
dimensionless parameter allows us to solve the non-linear field
equations for $B(t)$ pertubatively.  For a consistent expansion,
we must also specify how the sources of gravity --- the
cosmological constant, scalar field and Casimir contribution ---
scale with $\epsilon$.  In the solutions we have found, $\Lambda
M^{-2}$, $\frac{1}{4}\dot\phi^2 M^{-4}$, and $C M^{-4}$, are all
of order $\epsilon^2$.  In the following, we shall not write
factors of the Planck mass explicitly, absorbing $M^{-2}$ factors
into $\dot\phi^2$, $C$, $\rho$, $p$, etc.

At linear order in $\epsilon$ the field equations reduce to one
equation,
\begin{equation}
\frac{d^2B}{ dt^2} + 2\nu \frac{d^4B}{ dt^4} + \sum\limits_{k=3}^\infty
\sum\limits_{i} \nu_{k,i} \frac{d^{2k}B}{dt^{2k}} = 0\,,
\label{genosc}
\end{equation}
where $\nu = 3a+b+c$ and the $\nu_{k,i}$ are various combinations
of coefficients of terms higher order in $R$.  This equation has
sinusoidal solutions $B(t) = \epsilon \cos(\omega t)$ as long as
the polynomial
\begin{equation}
- \omega^2 + 2\nu\, \omega^4 + \sum\limits_{k=3}^\infty
\sum\limits_i (-1)^k \nu_{k,i} {\omega }^{2k}  = 0
\end{equation}
has at least one real root.  Thus for a range of parameters in
the original action --- so that no fine tuning is needed --- such
solutions exist.  The non-linearity of the field equations
appears in the next order terms in the expansion where the
amplitude $\epsilon$ is determined in terms of the cosmological
constant and other parameters in the action.

This behavior indicates that, at least in the small amplitude
limit, periodic metrics of the form~(\ref{Bmetric}) exist to an
arbitrary order in the derivative expansion.  This derivative
expansion and its associated solutions should arise from a
general theory as we near the Planck scale. \pagebreak[3]

To obtain more explicit results we shall hereafter drop the terms
in the action of order $R^3$ and higher. It should be clear that
the existence of the solutions does not depend on this
truncation.  While the quantitative results certainly are
sensitive to the truncation, the qualitative picture should not
be altered.  Since we are not expanding in powers of small
derivatives, the higher derivative terms in the equations are
just as important as lower derivative terms so that the usual
procedure of using equations of motion at low order in a
derivative expansion to simplify the analysis at higher order
does not apply here.

After the truncation we find
\begin{equation}
\omega  = \frac{1}{\sqrt{2\nu}}
\end{equation}
so that $\nu$ should satisfy the constraint $\nu > 0$.  When we
extend to order $\epsilon^2$, the analogue of~(\ref{genosc}) is a
non-linear differential equation whose solution is
\begin{equation}
B(t) = \epsilon \cos(\omega t)
- \frac{3}{16} \epsilon^2 \cos(2\omega t)\,,
\end{equation}
with
\begin{equation}
\epsilon^2  = - \frac{16}{ 3} \Lambda \nu\,,
\qquad \mbox{and} \qquad \dot\phi^2+4C = 4\Lambda\,.
\end{equation}
This result shows the importance of the cosmological constant for
the existence of an oscillating solution since the amplitude is
proportional to $\sqrt{-\Lambda}$.  Since $\nu>0$ we also see
that $\Lambda<0$.  Expanding to order $\epsilon^3$ we have
\begin{eqnarray}
B(t) &=& \epsilon \cos(\omega t) - \frac{3}{16} \epsilon^2\cos(2\omega t)
+ \frac{13}{ 256} \epsilon^3 \cos(3\omega t)\,, \nonumber\\
\omega &=& (2\nu)^{-1/2} \left( 1 - \frac{45}{64} \epsilon^2
\right), \nonumber\\
\dot\phi^2 + 4C &=& 4\Lambda
\left(1 - 3 \epsilon \cos(\omega t) \right).
\end{eqnarray}\looseness=-1
Only at this order does a time dependence of $\dot\phi$ appear,
showing the necessity of a dynamical field in addition to
gravity.  Note that the correction to the frequency is of order
$\epsilon^2$ rather than $\epsilon$. A numerical
analysis~\cite{periodic} of the full equations shows that there
really is an exact solution that is being approximated here and
such a solution exists as long as $0 < -\nu\Lambda < {1}/{24}$.

We see the first instance of a generic feature of oscillating
metric solutions, that the square of the oscillation amplitude
$\epsilon$ is proportional to the cosmological constant.  The $C$
parameter introduces a puzzle, as can be seen by considering the
leading order $\epsilon^2$ contributions to the energy-momentum
tensor,
\begin{eqnarray}
\rho &=&  \Lambda + \frac{1}{4} \dot\phi^2 + C = 2\Lambda < 0\,,
\nonumber\\
p &=& - \Lambda + \frac{1}{4} \dot\phi^2 + C = 0\,.
\end{eqnarray}
Such a density and pressure violates various positive energy
conditions~\cite{wald} so that it is difficult to see how the
required negative $C$ can arise.  While this apparent difficulty
might result from our truncation of the action beyond the $R^2$
order, we do not pursue this possibility here.

\subsection{Oscillating metrics in $4+1$ dimensions}
By extending to $4+1$ dimensions, we now have the possibility
that the behavior of the oscillations in the extra dimension
differs from that in the large dimensions,
\begin{equation}
ds^2 = - dt^2 + e^{B(t)} \delta_{ij} dx^i dx^j + e^{C(t)} dy^2\,.
\label{BCmetric}
\end{equation}
At first order in $\epsilon_t$, using\footnote{We introduce
subscripts to distinguish $\epsilon_t$ from $\epsilon_y$ used in
the next subsection.}
\begin{equation}
B(t) = \epsilon_t \cos(\omega_t t)\,, \qquad
C(t) = \epsilon_t \eta \cos(\omega_t t)
\end{equation}
two solutions to the field equations exist,
\begin{equation}
\eta = 1\,, \qquad \omega_t = \sqrt{\frac{3}{\mu}}\,,
\qquad \mbox{and}\qquad
\eta = -3\,, \qquad \omega_t = \frac{1}{\sqrt{3\mu-16\nu}}\,.
\end{equation}
We have chosen new linear combinations of the coefficients of the
$R^2$ terms in the five-dimensional action,
\begin{eqnarray}
\mu &=& 16a+5b+4c\,, \nonumber\\
\nu &=& 3a+b+c\,, \nonumber \\
\lambda &=& 5a+b+\frac{1}{2} c\,.
\end{eqnarray}
$\nu$ is the coefficient of the squared Weyl term.  The
coefficient of the Gauss-Bonnet term, $\lambda$, does not appear
until order $\epsilon_t^3$.

The first solution is completely analogous to the previous case
and it leads to the same problem. Therefore let us consider the
second solution and continue the expansion to order~$\epsilon_t^3$,
\begin{eqnarray}
B(t) &=& \epsilon_t \cos(\omega_t t) + b_2 \epsilon_t^2
\cos(2\omega_t t) + b_3 \epsilon_t^3
\cos(3\omega_t t)\,, \nonumber\\
C(t) &=& \epsilon_t \eta \cos(\omega_t t) + c_2 \epsilon_t^2
\cos(2\omega_t t) + c_3 \epsilon_t^3 \cos(3\omega_t t)\,, \nonumber \\
\frac{1}{4} \dot{\phi}^2 &=& \Lambda\,, \label{BCthree}
\end{eqnarray}
where the frequency and and amplitudes are
\begin{eqnarray}
\omega_t &=& \frac{1}{\sqrt{3\mu-16\nu}}
\left( 1 + \omega_{t,02} \epsilon_t^2 \right),  \nonumber\\
\eta &=& -3 \left( 1 + \eta_3 \epsilon_t^2 \right), \nonumber \\
\epsilon_t^2 &=& \frac{4}{3} \Lambda (3\mu-16\nu)\,,\label{hh}
\end{eqnarray}
and where
\begin{equation}
b_2 = c_2 = - \frac{3}{4} \frac{8\nu -\mu}{ 48\nu -5\mu}
\label{btwoctwo}
\end{equation}
with $b_3$, $c_3$, $\omega_{t,02}$, and $\eta_3$ as given in the
appendix. Here $\Lambda$ denotes the five-dimensional
cosmological constant.  We see that $\dot\phi^2$ remains constant
at this order, unlike the previous case, although this behavior
does not continue to hold at higher orders.  More importantly,
from the results for $\omega_t$ and $\epsilon_t^2$ we see that
$3\mu-16\nu$ and $\Lambda$ must both be positive, and thus we get
a consistent result for $\dot\phi^2$ without any additional
Casimir contribution.  The pressure and density which support
this geometry are $\rho = 2\Lambda > 0$ and $p = p_y = 0$.  The
change of sign for the energy density has occurred through the
introduction of an extra dimension which oscillates with a
different amplitude than the other spatial dimensions.

We might hope that since the cosmological constant is related to
the amplitude of an oscillating field, it could relax to zero
with the decay of this field. But in this scenario the amplitude
is directly related to $\Lambda$, a fundamental constant in the
action, and cannot relax.  This observation suggests that a
better approach would be to find a family of metrics in which
some \emph{effective} four-dimensional cosmological constant,
still related to an oscillating field, exists.  In such a
scenario, this effective cosmological constant is not forbidden
from relaxing while the fundamental $\Lambda$ does not change.

\subsection{Metrics periodic in an extra dimension}
Before examining metrics with these properties, we describe the
case of a static $4+1$-dimensional universe where a periodic
metric produces a naturally compact extra dimension.  Consider a
metric of the form
\begin{equation}
ds^2 = e^{A(y)} \eta_{\mu\nu} dx^\mu dx^\nu + dy^2\,.
\label{Ametric}
\end{equation}
where $A(y)$ is periodic and $y$ coordinates separated by one
period are identified.  In~\cite{periodic}, numerical solutions
of this form were studied in detail. In this scenario the size of
the compact space, the period, is dynamically determined, and
there is no radion stabilization problem. Rescaling
$dy^2\rightarrow \kappa^2 dy^2$ has no physical significance since
it would result in the period of $A(y)$ being scaled by
$\kappa^{-1}$.  The physical size of the compact space, the
product of scale factor and range of $y$, remains the same.

With a scalar field $\phi(y)$ we consider a contribution to the
energy-momentum tensor of the form $\rho = - p = p_y =
\frac{1}{4} {\phi'}^2 + C$.  We again introduce a constant $C$ as
a possible Casimir effect, and shall return to its role later.
The scalar field $\phi(y)$ is also compact, with the solution
determining the range over which it varies.  Thus we have a
dynamical generation of a compact internal space, in parallel
with the dynamical generation of a compact fifth dimension.  No
fine-tuning of parameters in the action is required for
either~\cite{periodic}.

The solution to order $\epsilon_y^3$ for the metric and the
scalar field is
\begin{eqnarray}
A(y) &=& \epsilon_y \cos(\omega_y y)- \frac{1}{4}
\epsilon_y^2 \cos(2\omega_y y) + \frac{1}{144}
\frac{13\mu-\lambda}{\mu} \epsilon_y^3 \cos(3\omega_y y)\,,\nonumber\\
{\phi'}^2 + 4C &=& -4 \Lambda \left( 1 - 4 \epsilon_y \cos(\omega_y y)
\right), \label{b}
\end{eqnarray}
where the frequency and the amplitude are given by
\begin{equation}
\omega_y = \sqrt{\frac{-3}{\mu}}
\left(1 + \frac{1}{4} \frac{\lambda-5\mu}{\mu} \epsilon_y^2 \right),
\qquad \epsilon_y^2 = - \frac{4}{9} \Lambda \mu\,.
\end{equation}\pagebreak[3]
The expressions for $\omega_y$ and $\epsilon_y^2$ indicate that
$\mu$ must be negative and $\Lambda$ must be positive.  The
leading contributions to the energy-momentum tensor yield $\rho =
p = 0$,  $p_y = -2\Lambda$.  Thus although the action has a
positive five dimensional $\Lambda$, the \emph{effective}
four-dimensional  cosmological constant vanishes.

Not surprisingly, it is also true that there are other solutions
in which this effective cosmological constant is
nonvanishing~\cite{warpedRS}. The apparent cosmological constant
seen by a low-energy observer is adjustable, depending upon the
warping of the fifth dimension. The problem here is that we don't
know why or how the system would relax to the flat low-energy
solution.

It is unclear whether a negative pressure in a fifth dimension
should be considered unphysical, but it is less of a concern for
the following reason. Even though a small five-dimensional
$\Lambda$ is convenient for our expansion, there is no reason
that it should be small in Planck units.  When $\Lambda$ is not
small and a higher derivative scalar term is included there are
periodic solutions that do not require the Casimir contribution,
$C$~\cite{periodic}.  The required energy-momentum tensor that
supports this geometry, which is in that case strongly $y$
dependent, arises explicitly from the scalar field configuration.

\section{Oscillating metrics with a warped extra dimension}\label{sec:three}

A universe with only a rapid time oscillation encounters the same
difficulty that occurs for a de Sitter solution --- the amplitude
of the former and the expansion rate of the latter are both
proportional to $\sqrt{\Lambda}$ so that no direct method for the
relaxation of either feature is available and we must simply
assume from the beginning an exponentially small size for
$\Lambda$ in its natural units.  In the static, warped scenarios
described in the previous subsection we do not need to make any
such fine-tuning of $\Lambda$ to obtain a universe that appears
approximately flat at low energies.  However, this solution
apparently involves tuning the initial conditions so that no
effective, low-energy cosmological constant is present.  In this
section we shall study metrics that depend periodically on both
time and the extra dimension.  The cosmological constant then
serves partially to drive the oscillations and partially to
support the compact dimension.  The oscillations are thereby no
longer forbidden from damping away since their amplitude is given
by a free parameter of the solution and not~$\sqrt{\Lambda}$.

Generalizing the metrics~(\ref{BCmetric}) and~(\ref{Ametric}) to
first order in the amplitudes in the two oscillations,
$\epsilon_t$ and $\epsilon_y$, we consider a metric of the form
\begin{equation}
ds^2 = - e^{A(y)} dt^2 + e^{A(y)+B(t)} \delta_{ij}
dx^i dx^j + e^{C(t)} dy^2\,, \label{g}
\end{equation}
with
\begin{equation}
A(y) = \epsilon_y \cos(\omega_y y)\,, \qquad B(t) = \epsilon_t
\cos(\omega_t t)\,, \qquad C(t) = - 3\epsilon_{t} \cos(\omega_t t)\,.
\end{equation}
We have already seen that
\begin{equation}
\omega_y = \sqrt{\frac{-3}{\mu}}\,, \qquad \mbox{and} \qquad
\omega_t = \frac{1}{\sqrt{3\mu-16\nu}}\,.\label{u}
\end{equation}
The range of $y$ is determined by $\omega_y$, but the size of the
compact space oscillates in time with frequency $\omega_t$.

At next order in $\epsilon_t$ and $\epsilon_y$ the nonlinear
field equations induce a mixing between the $t$ and $y$ dependent
terms, and thus the metric assumes a more complicated form,
\begin{equation}
ds^2 = - e^{A(y)+E(y,t)} dt^2+ e^{A(y)+B(t)+F(y,t)}
\delta_{ij} dx^i dx^j + e^{C(t)+G(y,t)} dy^2\,.
\label{ABCmetric}
\end{equation}
At second order we find
\begin{eqnarray}
A(y) &=& \epsilon_y \cos(\omega_y y)- \frac{1}{ 4}
\epsilon_y^2 \cos(2\omega_y y)  \,, \nonumber \\
B(t) &=& \epsilon_t \cos(\omega_t t)+ b_2 \epsilon_t^2
\cos(2\omega_t t)\,, \nonumber \\
C(t) &=& - 3\epsilon_t \cos(\omega_t t)
+ c_2 \epsilon_t^2 \cos(2\omega_t t)\,, \nonumber \\
E(y,t) &=& e_{11} \epsilon_y \epsilon_t \cos(\omega_y y)
\cos(\omega_t t)\,, \nonumber \\
F(y,t) &=& f_{11} \epsilon_y \epsilon_t \cos(\omega_y y)
\cos(\omega_t t)\,, \nonumber \\
G(y,t) &=& g_{11} \epsilon_y \epsilon_t \cos(\omega_y y)
\cos(\omega_t t)\,. \label{Gtwo}
\end{eqnarray}
$b_2$, and $c_2$ are as before, $f_{11}$ and $g_{11}$ are given
in the appendix, and $e_{11}$ remains undetermined.\footnote{The
value of $e_{11}$ does not affect five-dimensional curvature
invariants at this order, and its value, given in the appendix,
is determined at third order in the expansion.} We find that
\begin{eqnarray}
\epsilon_y^2 &=& \frac{4}{ 3} \frac{\Lambda}{\omega_y^2}
- \frac{\omega_t^2}{\omega_y^2} \epsilon_t^2\,, \nonumber \\
{\phi'}^2 + 4C &=& - 4\Lambda + 3\omega_t^2\epsilon_t^2\,, \nonumber \\
\dot\phi^2 &=& 3\omega_t^2\epsilon_t^2\,,\label{a}
\end{eqnarray}
where $\omega_y$ and $\omega_t$ are the first order values in~(\ref{u}).

The significant feature of this solution is that the cosmological
constant is related to \emph{both} amplitudes --- we can freely
vary $\epsilon_t$, for example, with a fixed value of $\Lambda$,
and still find a solution since $\epsilon_y$ can compensate for
the changes in the time oscillations.  To emphasize the
similarity of this component of the solution and the behavior we
saw in the previous section, where a cosmological constant drove
an oscillating metric in $3+1$ dimensions, we introduce a new
positive constant, $\Lambda_\mathrm{osc}$,
\begin{equation}
\Lambda_\mathrm{osc} \equiv \Lambda + C + \frac{1}{4} {\phi'}^2
= \frac{3}{ 4} \omega_t^2\epsilon_t^2\,.
\label{d}
\end{equation}
Note that $\Lambda_\mathrm{osc}$ only parameterizes the size of
the time oscillations in the metric and is not fixed by any
quantity in the action.  In the limit $\Lambda_\mathrm{osc}\to
0$, we recover the static metric of eq.~(\ref{Ametric}).

In terms of this new parameter, the density and pressure which
support this geometry have the form
\begin{eqnarray}
\rho &=& \Lambda_\mathrm{osc} + \frac{1}{4} \dot\phi^2
= 2\Lambda_\mathrm{osc} > 0\,, \nonumber \\
p &=& - \Lambda_\mathrm{osc} + \frac{1}{ 4} \dot\phi^2 = 0
\label{ABCrho}
\end{eqnarray}
in the large directions.  $\Lambda_\mathrm{osc}$ appears here
rather than the five-dimensional $\Lambda$.  The latter drives
the warping of the fifth dimension and appears in the pressure in
the extra dimension, $p_y = - 2\Lambda + 2\Lambda_\mathrm{osc}$.
Note that since the action contains terms quadratic in the
curvature tensors, the energy-momentum tensor is not
conventionally related to the Einstein tensor, $G_{\mu\nu}$.  For
example, when averaging over the fifth dimension we find
\begin{equation}
G_{tt} = - 2\Lambda_\mathrm{osc} \sin^2(\omega_t t)\,.
\end{equation}
This equation could perhaps be interpreted as an effective energy
density obtained by combining the contributions from the $R^2$
terms together with the actual contributions to the
energy-momentum tensor.

In principle, we can continue the small amplitude expansion to
higher orders although the expressions for the new coefficients
quickly become quite lengthy. At third order, the expressions for
the frequencies receive small corrections,
\begin{eqnarray}
\omega_t &=& \frac{1}{\sqrt{3\mu-16\nu}}
\left(1 + \omega_{t,02} \epsilon_t^2 + \omega_{t,20}\epsilon_y^2
\right), \nonumber\\
\omega_y &=&  \sqrt{\frac{-3}{\mu}}
\left( 1 + \frac{1}{ 4} \frac{\lambda-5\mu}{\mu} \epsilon_y^2 +
\omega_{y,02}\, \epsilon_t^2 \right).\label{FGomega}
\end{eqnarray}
The third order corrections involving terms with purely $t$ or
$y$ dependence are as in the previous sections.  $\eta$ is as
in~(\ref{hh}), the right side of the second equation in~(\ref{a})
is multiplied by the same $(1 - 4
\epsilon_y \cos(\omega_y y) )$ factor appearing in~(\ref{b}), and the third
equation in~(\ref{a}) is not corrected.\footnote{These results
are consistent with $\phi(t,y)=\phi_1(t)+\phi_2(y)$, but
alternatively there could be two independent scalar fields with
dependence only on $t$ and $y$, respectively.} The third order
corrections involving a product of a $t$ or a $y$ dependence are
contained in the following expansions:
\begin{eqnarray}
F(y,t) &=& f_{11} \epsilon_y\epsilon_t
\cos(\omega_y y)\cos(\omega_t t) + f_{21}
\epsilon_y^2\epsilon_t\cos(2\omega_y y)\cos(\omega_t t)
 \nonumber \\
&& +\, f_{12} \epsilon_y\epsilon_t^2 \cos(\omega_y y) \cos(2\omega_t t)\,,
\nonumber \\
G(y,t) &=& g_{11} \epsilon_y\epsilon_t \cos(\omega_y y) \cos(\omega_t t)
+ g_{21} \epsilon_y^2\epsilon_t \cos(2\omega_y y) \cos(\omega_t t)
\nonumber \\
&& +\, g_{12} \epsilon_y\epsilon_t^2 \cos(\omega_y y) \cos(2\omega_t t)
+ g_{21}^\prime \epsilon_y^2\epsilon_t \cos(\omega_t t)\,.\label{Gthree}
\end{eqnarray}
The expressions for the coefficients in~(\ref{FGomega})
and~(\ref{Gthree}) are given in the appendix.

\section{Evolution and cosmology}\label{sec:four}

The previous section introduced a class of metrics that are
periodic in both time and an extra dimension in a universe with a
non-finely tuned cosmological constant.  The existence of a
small, compact dimension will not be apparent to observers who
can only probe length scales much larger than $\sqrt{-\mu}$.  In
contrast, the existence of a Planck-frequency oscillation would
not be compatible with current observations~\cite{horowitz}
unless its amplitude is extremely small~\cite{fermions}.  Since
the time oscillations are driven by a positive contribution to
the energy density~(\ref{ABCrho}) the oscillating metric
in~(\ref{ABCmetric}) would seem to be unstable against relaxing
toward a metric~(\ref{Ametric}) which is flat in the large
dimensions.  To investigate the evolution of the oscillation
amplitude and its dynamical coupling to the cosmological
expansion would require the addition of some time dependence to
the amplitudes, $\epsilon_t\to\epsilon_t(t)$ and
$\epsilon_y\to\epsilon_y(t)$.  However, analyzing this more
general time dependence becomes intractable for the above
analytic expansion.  In this section we shall restrict therefore
to showing how this picture differs from a Kaluza-Klein model
with a flat extra dimension and to understanding how a rapid
oscillation of \emph{constant} amplitude is  incompatible with a
cosmological expansion.

When the large dimensions evolve with a Robertson-Walker
expansion --- assuming the oscillations have decayed --- this
scenario behaves very differently from a standard Kaluza-Klein
cosmology in which two time dependent scale factors govern the
evolution of the size of the large spatial dimensions and the
compact space, respectively.  In this latter case, one must ensure
that the resulting time evolution of physical constants after
integrating out the extra dimensions is sufficiently
small~\cite{kolb}. In contrast, as described in
section~\ref{sec:three}, the second scale factor in the warped
scenario is not an independent dynamical quantity since the size
of the compact space is already determined.  This property makes
a standard cosmological evolution more natural.  For example a
power law scale factor $a(t)=t^\alpha$ for the three large
spatial dimensions is determined as usual by
\begin{equation}
\alpha =  \frac{2}{3} \frac{1}{1+w}\,,
\end{equation}
where $p=w\rho$.  For a given $w$ the time dependence of $\rho$
and $p$ is determined, as is the matter contribution to the
five-dimensional pressure,
\begin{equation}
\left. p_y \right|_\mathrm{matter} = \frac{1}{2} (3w-1) \rho\,.
\label{i}
\end{equation}
This result follows directly from the five-dimensional Einstein
equation given that the compact dimension has fixed size.  When
the variation in the non-compact coordinates is small compared to
the size of the compact dimension, the contribution of ordinary
matter and radiation fields to the pressure in the extra
dimension will also be small compared to the order $\epsilon_y$
contribution to $p_y$; the latter is $-2\Lambda$ which drives the
warped compactification in the first place.

If we attempt to introduce cosmological expansion with a scale
factor $a(t)$ along with a rapid oscillation of constant
amplitude,
\begin{equation}
ds^2 = e^{A(y)} \left( - dt^2 + e^{\epsilon_t
\cos(\omega_t t) + \cdots} a^2(t) \delta_{ij} dx^i
dx^j \right) + e^{-3 \epsilon_t \cos(\omega_t t) +\cdots} dy^2\,,
\label{ABCfmetric}
\end{equation}
we should apparently choose $B(t) = \epsilon_t \cos(\omega_t t) +
f(t)$, where $f(t)\equiv 2\ln(a(t))$.  However, inserting this
metric into the leading small amplitude field equations produces
crossterms of the form $\dot f(t) \sin(\omega_t t)$ which do not
cancel.  Moreover, they cannot be canceled by ordinary matter
effects because we are assuming that the latter is a long
wavelength effect.  We can eliminate such terms and thereby find
solutions for the \emph{gravitational} field equations if we
include the following dependence on $f(t)$ in $B(t)$ and $C(t)$:
\begin{eqnarray}
B(t) &=& \epsilon_t \cos(\omega_t t) + f(t)
+ b_4 \epsilon_t \dot f(t) \omega_t^{-1} f\sin(\omega_t t)\,,\nonumber\\
C(t) &=& -3 \epsilon_t \cos(\omega_t t) + c_4 \epsilon_t \dot f(t)
\omega_t^{-1} \sin(\omega_t t)\,, \nonumber \\
\dot\phi^2(t) &=& 3\omega_t^2\epsilon_t^2 + 6 \epsilon_ t \dot f(t)
\omega_t \sin(\omega_t t) \label{q}
\end{eqnarray}
with
\begin{equation}
b_4 = \frac{2}{3} \frac{\omega_t^2}{\omega_y^2} \frac{1}{\epsilon_y^2}
 + \frac{3\omega_y^2-\omega_t^2}{2\omega_y^2 + 2\omega_t^2}\,,
\qquad c_4 = - 2 \frac{\omega_t^2}{\omega_y^2} \frac{1}{\epsilon_y^2}\,.
\end{equation}
Defining $\dot f(t) \approx \epsilon_f M$, the size of the $b_4$
and $c_4$ terms are of order
$\epsilon_t\epsilon_f\epsilon_y^{-2}$ and the expansion appears
to make sense for $B(t)$ and $C(t)$ in the regime of interest,
$\epsilon_t \approx \epsilon_f \ll \epsilon_y$.

The incompatibility of the cosmological expansion with the
oscillations arises in the expression for $\dot\phi^2(t)$, where
we see that the correction is of the same order as the first
term.  The presence of this term does not allow the \emph{scalar}
field equation for $\phi(t)$ to be satisfied.  This difficulty is
presumably related to the fact that the small amplitude expansion
does not account for a varying amplitude for the oscillations.
This investigation is continued numerically in $3+1$ dimensions
in~\cite{holdom}.

\section{Conclusions}\label{sec:five}

In the very early universe, when higher order curvature
invariants become important in the action, the presence of a
cosmological constant can produce a rapidly oscillating metric
rather than a de Sitter expansion.  In a $3+1$-dimensional
universe or a universe with a flat extra dimension, since the
cosmological constant is directly related to the amplitude of
these oscillations, no method for damping these oscillations is
available.  In a universe with a warped extra dimension, the
cosmological constant can play a dual role both supporting a
compact extra dimension and, as before, driving oscillations.
Since the relative amount which the cosmological constant
supports these two periodic behaviors is a free parameter, the
oscillations can decay through their coupling to other fields
present.  In the resulting approximately static universe, the
cosmological constant solely supports the compact dimension and
so does not need to be small to produce an effectively flat
$3+1$-dimensional universe in the low energy limit.

We have used this scenario as a illustration of how extra
dimensions and a more general response to including a $\Lambda$
in the action can evade the usual theorems forbidding the
relaxation of a cosmological constant --- since in fact $\Lambda$
does not change and instead only an effective parameter damps
away.  Nevertheless, significantly more work is needed to apply
this approach to understanding the cosmological constant problem.

The detailed results of this paper have relied on a truncation of
the higher derivative terms to the $R^2$ order. Consideration of
higher order terms may open up even more possibilities, but they
should not qualitatively change the possibilities we have
discussed.  These possibilities could also exist for the true
underlying theory for gravity.

The time-dependent oscillations have appeared as solutions
disjoint from the usual exponentially expanding de Sitter
solutions, since both types of time dependence cannot coexist in
our version of the small amplitude expansion.  It may be that a
more general type of solution exists for which pure oscillations
and a pure cosmological expansion exist as limiting cases.  This
possibility is pursued in $3+1$ dimensions in~\cite{holdom}.  A
nontrivial coupling between the amplitude of oscillations and the
large scale expansion then leaves open the possibility that
oscillations could exist even at late times.

\acknowledgments

This work was supported in part by the Natural Sciences and
Engineering Research Council of Canada.

\appendix

\section{Higher order coefficients}

In this appendix, we list some of the more cumbersome
coefficients of the higher order terms in the various small
amplitude expansions.  In the solution to the five-dimensional
metric with only time dependence~(\ref{BCthree}) we have
\begin{eqnarray}
b_3 &=& \frac{-15\lambda\mu^2 + 239\mu\lambda\nu +
30\mu^3 - 1018\nu\mu^2 + 9780\mu\nu^2 - 912\lambda\nu^2 - 27648\nu^3 }{
432\nu ( 224\nu\mu - 15\mu^2 - 768\nu^2 )}\,, \qquad\qquad\\[3pt]
c_3 &=& \frac{30\mu^3 - 15\lambda\mu^2 + 722\nu\mu^2 + 179\mu\lambda\nu
- 15132\mu\nu^2 - 336\lambda\nu^2 + 55296 \nu^3 }{
432\nu ( 224\nu\mu - 15\mu^2 - 768\nu^2 )}\,, \\[3pt]
\omega_{t,02} &=& \frac{ 97\mu^2 - 5\lambda\mu - 1452\nu\mu + 4992\nu^2
+ 48\lambda\nu} {4 ( 224\nu\mu - 15 \mu^2 - 768\nu^2 )}\,,  \\[3pt]
\eta_3 &=& \frac{-\lambda + 2\mu - 9\nu }{ 6(\mu-6\nu)}\,.
\end{eqnarray}
In the metrics with both $t$ and $y$ dependence, at second order,
the coefficients of the mixing terms~(\ref{Gtwo}) are
\begin{eqnarray}
f_{11} &=& \frac{- 13824\nu^2 + 5124\mu\nu -
32\lambda\mu + 192\lambda\nu - 477\mu^2}{ 6\mu (3\mu-16\nu)}\,, \\[3pt]
g_{11} &=& - 3 e_2 \frac{(3\mu-16\nu)}{\mu}
+ \frac{221184\nu^3 + 24648\mu^2\nu - 128064\mu\nu^2}{2\mu^2
(3\mu-16\nu)} \nonumber\\[3pt]
&& -\, \frac{3072\lambda\nu^2 + 112\lambda\mu^2 +
1575\mu^3 - 1152\nu\lambda\mu }{2\mu^2 (3\mu-16\nu)}\,.
\end{eqnarray}
The third order corrections~(\ref{FGomega}) and~(\ref{Gthree}) have
the following coefficients.
\begin{eqnarray}
\omega_{y,02} &=& \left( - 1536\nu\lambda^2 - 80448\nu
\lambda\mu - 783558\mu^2\nu + 256\mu\lambda^2 + 7344\lambda\mu^2 +
221184\lambda\nu^2 \right. \nonumber\\[3pt]
&& \left.\hphantom{(}\! + \, 4326912\mu\nu^2 - 7962624\nu^3 + 47277\mu^3 \right)
\Big/ \left( 288\mu (3\mu-16\nu)^2\right),  \\[3pt]
\omega_{t,20} &=& - \left( 50085\mu^3 - 814086\mu^2\nu +
7344\lambda\mu^2 - 80448\nu\lambda\mu + 256\mu\lambda^2 +
4409856\mu\nu^2 \right. \nonumber\\[3pt]
&& \left.\hphantom{-(}\! -\, 1536\nu\lambda^2 + 221184\lambda\nu^2 -
7962624\nu^3\right)\Big/ \left(96\mu^2(3\mu-16\nu)\right), \\[3pt]
e_{11} &=& -\left( 15925248\nu^4 - 10215936\mu\nu^3 +
207360\mu\lambda\nu^2 + 6165\mu^4 + 4864\mu^3\lambda 
\right.\nonumber \\[3pt]
&& \hphantom{-(}\!+\, 256\mu^2\lambda^2 -
218382\mu^3\nu - 1536\mu\nu\lambda^2 -
221184\lambda\nu^3 + 2344032\mu^2\nu^2 \nonumber \\[2pt]
&&\hphantom{-(}\! -\,\left. 56640\mu^2\lambda\nu \right) \Big/
\left( 72\mu (\mu-6\nu) (3\mu-16\nu)^2 \right),  \\[2pt]
f_{21} &=& - \left( 285273\mu^5 - 859963392\nu^5 +
135168\nu\mu^2\lambda^2 - 741888\mu\nu^2\lambda^2 \right. \nonumber \\[2pt]
&&\hphantom{-(}\! +\, 65992320\mu\lambda\nu^3 - 7035288\mu^4\nu +
2479416\mu^3\lambda\nu + 1354752\nu^3\lambda^2  \\[2pt]
&&\hphantom{-(}\! -\, 8192\mu^3\lambda^2 - 120780\mu^4
\lambda + 69600906\mu^3\nu^2 -
19152288\mu^2\nu^2\lambda - 85598208\lambda\nu^4 \nonumber \\[2pt]
&&\hphantom{-(}\! +\, \left. 860087808\mu\nu^4 - 345400308\mu^2\nu^3
\right) \Big/ \left( 144\mu(\mu-6\nu) (3\mu-16\nu)^2 (7\mu-36\nu)
\right), \nonumber \\[2pt]
f_{12} &=& \left(-36720677376\mu^2\nu^5 + 199247877\mu^5\nu^2 -
11515515\mu^6\nu + 640\mu^5\lambda^2 \right. \nonumber \\[2pt]
&& \hphantom{(}\! +\, 29098\mu^6\lambda + 282591\mu^7 +
10818446400\mu^3\nu^4 - 55037657088\nu^7 \nonumber \\[2pt]
&&\hphantom{(}\! +\, 1528823808\nu^6\lambda - 10616832\nu^5\lambda^2 -
25344\mu^4\nu\lambda^2 - 1901363418\mu^4\nu^3 -\nonumber \\[2pt]
&&\hphantom{(}\! -\, 1132452\mu^5\lambda\nu - 146614464\mu^3\lambda\nu^3 +
17867184\mu^4\lambda\nu^2 + 68860772352\mu\nu^6 \nonumber \\[2pt]
&&\hphantom{(}\!-\, 1569964032\mu\nu^5\lambda - 2566656\mu^2\nu^3\lambda^2 +
662611968\mu^2\nu^4\lambda + 372480\mu^3\nu^2\lambda^2 \nonumber \\[2pt]
&&\hphantom{(}\! \left. +\, 8404992\nu^4\mu\lambda^2 \right) \Big/
\left( 72\mu^2 (\mu-6\nu) (3\mu-16\nu)^2 (\mu-4\nu)(5\mu-48\nu) \right), \\[2pt]
g_{21} &=& - \left(7735005\mu^6 + 110075314176\nu^6 +
131254272\nu^3\lambda^2\mu + 4668928\mu^3\nu\lambda^2 \right. \nonumber\\[2pt]
&&\hphantom{(+}\!+\, 4215635712\lambda\nu^3\mu^2 - 18316369656\mu^3\nu^3
\!-\! 10733617152\nu^4\lambda\mu \!+\! 81818128\mu^4\lambda\nu \nonumber \\[2pt]
&&\hphantom{(+} \!- 37174272\nu^2\lambda^2\mu^2 \!+\! 2793157068
\mu^4\nu^2 \!-\! 3233928\mu^5\lambda \!-\! 227538114\mu^5\nu 
\qquad\ \ \  \\[2pt]
&&\hphantom{(+} \!+\, 67677175296\nu^4\mu^2 + 10956570624\nu^5\lambda -
133596905472 \nu^5\mu - 219392\mu^4\lambda^2  \nonumber \\[2pt]
&& \left.\hphantom{(+}\! -\, 173408256\nu^4\lambda^2 - 829677504\mu^3
\lambda\nu^2 \right)\! \Big/ \!\left( 96\mu^2(\mu-6\nu)(3\mu-16\nu)^2
(7\mu-36\nu) \right), \nonumber \\[2pt]
g_{12} &=& \left( 2332845\mu^8 - 5361075486720\mu\nu^7 -
97844723712\lambda\nu^7 + 70400\mu^6\lambda^2 \right. \nonumber \\[2pt]
&&\hphantom{(}\! +\, 3522410053632\nu^8 + 277363541376\mu^4\nu^4 -
37624542552\mu^5\nu^3 - \nonumber \\[2pt]
&&\hphantom{(}\! -\, 132976434\mu^7\nu - 1262573733888\mu^3\nu^5 +
1351640\mu^7\lambda \nonumber \\[2pt]
&&\hphantom{(}\! +\, 3486633984000\mu^2\nu^6 + 3049013340\mu^6\nu^2 -
45487152\mu^6\lambda\nu \nonumber \\[2pt]
&&\hphantom{(}\! +\, 20788224\mu^4\nu^2\lambda^2 - 721944576\mu
\nu^5\lambda^2 + 679477248\nu^6\lambda^2 \nonumber \\[2pt]
&&\hphantom{(}\!+\, 646135872\mu^5\lambda\nu^2 -
1927680\mu^5\nu\lambda^2 - 5107325184\mu^4\lambda\nu^3 \nonumber \\[2pt]
&&\hphantom{(}\!+\, 126977310720\mu\nu^6\lambda +
376012800\nu^4\mu^2\lambda^2 + 24644653056\mu^3\nu^4\lambda \nonumber \\[2pt]
&& \left. \hphantom{(}\!-\, 116299776\mu^3\nu^3\lambda^2 -
73583493120\mu^2\nu^5\lambda \right) \nonumber \\[2pt]
&&\hphantom{(}\! \Big/\, \left( 384\mu^3(\mu-4\nu)(5\mu-48\nu)
(\mu-6\nu)(3\mu-16\nu)^2\right), \\[2pt]
g_{21}^\prime &=& \left( 9243\mu^4 + 5308416\nu^4 - 3072\nu^2\lambda^2 +
1335540\mu^2\nu^2 - 88704\mu\lambda\nu^2 \right. \nonumber\\[2pt]
&&\hphantom{(}\! +\, 1280\mu\nu\lambda^2 + 18032\mu^2\lambda\nu -
4359168\mu\nu^3 - 1244\mu^3\lambda - 128\mu^2\lambda^2 \nonumber \\[2pt]
&&\hphantom{(} \left.\! -\, 181428\mu^3\nu +
147456\lambda\nu^3 \right) \Big/
\left( 4\mu^2(3\mu-16\nu)(\mu-6\nu)\right).
\end{eqnarray}

\end{document}